\documentclass[superscriptaddress,nofootinbib,11pt,floatfix]{revtex4-2}

\usepackage{epsfig}
\usepackage{calrsfs}
\usepackage{times}
\usepackage{psfrag}
\usepackage{bbold}
\makeatletter\AtBeginDocument{\let\@elt\relax}\makeatother
\usepackage[title]{appendix}
\usepackage{xr}

\usepackage{placeins}

\usepackage[utf8]{inputenc}
\usepackage{hyperref}
\usepackage{color}
\usepackage{graphicx}
\usepackage{comment}
\usepackage{siunitx}
\usepackage{amsmath}
\usepackage{floatrow}

\usepackage[font=small,labelfont=bf,justification=justified,format=plain]{caption}

\usepackage[dvipsnames]{xcolor}
\usepackage{tikz}
\usepackage{subfigure}
\usepackage{breakurl}

\begin{document}

\title{Global misinformation spillovers in the online vaccination debate \\ before and during COVID-19} 

\author{Jacopo Lenti}
\affiliation{ISI Foundation, via Chisola 5, 10126, Turin, Italy}
\affiliation{CENTAI, corso Inghilterra 3, 10138, Turin, Italy}
\affiliation{Sapienza University, P.le A. Moro 5, 00185, Rome, Italy}

\author{Kyriaki Kalimeri}
\affiliation{ISI Foundation, via Chisola 5, 10126, Turin, Italy}

\author{Andr\'{e} Panisson}
\affiliation{CENTAI, corso Inghilterra 3, 10138, Turin, Italy}

\author{Daniela Paolotti}
\affiliation{ISI Foundation, via Chisola 5, 10126, Turin, Italy}

\author{\\Michele Tizzani}
\affiliation{ISI Foundation, via Chisola 5, 10126, Turin, Italy}

\author{Yelena Mejova}
\thanks{Corresponding author: yelenamejova@acm.org}
\affiliation{ISI Foundation, via Chisola 5, 10126, Turin, Italy}

\author{Michele Starnini}
\thanks{Corresponding author: michele.starnini@gmail.com}
\affiliation{Departament de Fisica, Universitat Politecnica de Catalunya, Campus Nord, 08034 Barcelona, Spain}
\affiliation{CENTAI, corso Inghilterra 3, 10138, Turin, Italy}

\begin{abstract}
Anti-vaccination views pervade online social media, fueling distrust in scientific expertise and increasing vaccine-hesitant individuals.
While previous studies focused on specific countries, the COVID-19 pandemic brought the vaccination discourse worldwide, underpinning the need to tackle low-credible information flows on a global scale to design effective countermeasures.
Here, we leverage 316 million vaccine-related Twitter messages in 18 languages, from October 2019 to March 2021, to quantify misinformation flows between users exposed to anti-vaccination (no-vax) content.
We find that, during the pandemic, no-vax communities became more central in the country-specific debates and their cross-border connections strengthened, revealing a global Twitter anti-vaccination network. 
U.S. users are central in this network, while Russian users also become net exporters of misinformation during vaccination roll-out. 
Interestingly, we find that Twitter’s content moderation efforts, and in particular the suspension of users following the January 6th U.S. Capitol attack, had a worldwide impact in reducing misinformation spread about vaccines.
These findings may help public health institutions and social media platforms to mitigate the spread of health-related, low-credible information by revealing vulnerable online communities.
\end{abstract}

\maketitle

\section{Introduction}

The COVID-19 pandemic has brought vaccination from the purview of parents and the health-compromised to everyone in the public.
Restrictions around vaccination brought an additional potential to impact one's personal freedoms and the world economy, as well as one's health.
However, vaccination hesitancy continues to limit the impact of this highly effective intervention\footnote{\url{https://www.who.int/emergencies/ten-threats-to-global-health-in-2019}}: hundreds of thousands of lives were lost to COVID-19 that could have been prevented with vaccinations in U.S. alone \cite{amin2022mortality}.  

Vaccination hesitancy is a complex issue which has been associated with science denial \cite{browne2015going}, alternative health practices \cite{Kalimeri2019}, and belief in conspiracy theories \cite{jolley2014effects}.
Among the many factors contributing to vaccine hesitancy is the spread of misinformation, especially online \cite{burki2019vaccine, carrieri2019vaccine}.
{The impact of anti-vaccination content on Online Social Media (OSM) may be compounded by the so-called echo-chamber effect \cite{cinelli2021echo}, in which users have their beliefs reinforced through interactions with like-minded peers \cite{del2016echo,zollo2015emotional,garimella2018political}. }
Created by the interplay of (i) homophily between users' interactions and (ii) polarization of the debate, echo chambers arise from a combination of psychological tendencies of confirmation bias and selective exposure \cite{quattrociocchi2016echo,cinelli2020selective,bakshy2015exposure} together with algorithmic optimization for greater engagement at the cost of content diversity \cite{schmidt2017anatomy}.
Importantly, echo-chambers have also been found on OSM in the discussions around vaccination \cite{cossard2020falling, bello2017detecting, schmidt2018polarization, crupi2022echoes}.

Thus far, the scientific study of the debate around vaccination on OSM has focused on specific countries \cite{cossard2020falling, crupi2022echoes, faccin2022assessing,  motta2020right} or English-speaking users  \cite{schmidt2018polarization}.
Nevertheless, the COVID-19 pandemic brought the vaccination discourse to a global scale \cite{ng2021news}, creating a deluge of international news around the development and deployment of COVID-19 vaccines, including low-quality content and misinformation \cite{rovetta2020global}.
The danger of this \emph{`infodemic'} has been acknowledged in mid 2020 by the UN and WHO, which called for member states to develop and implement the necessary action plans \cite{who2020managing}.
Thus, it is imperative to understand the flows of anti-vaccine --- or \emph{no-vax} --- information not only nationally but internationally, in order to have a bird-eye view on the topic and inform effective communication campaigns.

To address this need, in this work we focus on the Twitter platform by leveraging 316 million 
tweets related to vaccines in 18 different languages from a pre-COVID era to April 2021 to quantify misinformation flows between users in no-vax communities across national borders, and identify which countries are central in the global vaccination debate.
To this aim, we first investigate (i) how polarized, in terms of echo chambers phenomenon, the vaccination debate is in different countries, over time, to identify users in no-vax communities and (ii) how susceptible, in terms of circulation of information, are these no-vax communities to low quality information.
We propose a flexible, language-neutral community detection approach, and combine it with human-in-the-loop expert knowledge to track polarization and echo chambers in different countries and time periods. 
We show that communities in which no-vax content is shared (i) increased in number during the pandemic, (ii) became less isolated in the national vaccination debate, and (iii) displayed much stronger cross-border connections than the rest of the users.
Alarmingly, users in these communities tend to heavily rely on low-credibility information sources and to spread it across national borders, resulting in international spillovers of misinformation through a global no-vax network.

\section{Related Works}


Vaccination deliberation on Twitter has been studied mainly in English and in the U.S. \cite{kennedy2011confidence,blankenship2018sentiment,meadows2019twitter}.
However, recently the platform has gained attention of researchers also focusing on European countries. 
Before the pandemic, analysis of the Dutch Twitter revealed an antagonistic relationship between an `anti-establishment' community and that of journalists and writers, reinforcing the `arrogance of the elite' worldview in the former~\cite{lutkenhaus2019mapping}.
In the Italian Twitter, the debate around vaccination revealed polarization in terms of retweets,
where vaccine skeptics often mention vaccine advocates (generally in attacks), while the advocates seem to ignore the skeptics altogether~\cite{cossard2020falling}.
Outside Europe, 
a randomized study in Indonesian Twitter 
showed the importance of celebrity endorsement in message engagement, and that the inclusion of the information source is associated with decreased propagation~\cite{alatas2019celebrities}.
The COVID-19 pandemic has spurred increased attention to the topic.
A recent examination of vaccine-critical actors on francophone Twitter found that their place in discussions on vaccines has remained relatively constant during the pandemic, compared to mainstream media~\cite{faccin2022assessing}.
Further, \cite{crupi2022echoes} have studied the Italian Twitter during the roll-out of the COVID vaccinations, showing greater engagement across vaccine supporting and hesitant communities in terms of mentions and in the similarity of the topics discussed. 

Attempts to study the flows of vaccination discussion across borders have thus far been limited to dyadic relationships, and to English.
A study of Canadian Twitter users found that the majority of misinformation circulating on Twitter that is shared by Canadian accounts is retweeted from U.S.-based accounts, and that increased exposure to U.S.-based information on Twitter is associated with an increased likelihood to post misinformation~\cite{bridgman2021infodemic}.
Beyond Twitter, \cite{ng2021news} have examined news articles about COVID from 20 countries, identifying the shift in narratives as pandemic occurred.
However, the data was limited to English language, and failed to capture local-language coverage.
Unlike the previous efforts, our study tracks vaccination debate in the native languages of numerous countries to systematically study the flow of information (and potential misinformation) across national boarders.


The most concerning aspect of the vaccination debates studied here is misinformation that may damage the confidence in the procedure.
Controlled exposure studies have shown that online misinformation---especially scientific-sounding one---negatively impacts vaccination intent in subjects in US, UK~\cite{loomba2021measuring}, and New Zealand~\cite{thaker2021exposure}.
A panel study of US Twitter users found that the risk of average users occasionally sharing misinformation was alarmingly high, despite social bots’ contribution to misinformation
sharing being ``surprisingly low''~\cite{teng2022characterizing}.
While some efforts have been made toward high-quality, manually-annotated datasets for identifying misinformation~\cite{weinzierl2021misinformation}, the quality of the cited URL domains is often used as a gauge of the quality of the tweet's content~\cite{muric2021covid,yang2021covid}. 
In this work, we use a similar approach by combining lists of low-credibility domains from several languages and countries.

Beyond content analysis, an important aspect of (mis)information spread is the network structure underlying such dynamical process. 
Echo-chambers in the Twitter debate around the impeachment of former Brazilian President Dilma Rousseff have been shown to alter the diffusion of information between the two sides, supporters and opponents of the impeachment \cite{cota2019quantifying}.
A similar methodology has been used to compare different topics across social media \cite{cinelli2021echo}, highlighting that Facebook shows a higher segregation of news consumption than Reddit.
Along the same research line, the random walk controversy score \cite{garimella2016quantifying} quantifies how controversial are topics discussed over a certain social network, as the probability for an average user to be exposed to information from their own side versus from the opposing side. 
While several works address the presence of echo-chambers on social media and their effect on information diffusion, little to no efforts have been devoted to understanding the echo-chambers effects within cross-border information spreading, which we examine in this study.

\section{Methodology} 

We use Twitter Streaming API to collect a multi-lingual dataset, which we geolocate using the GeoNames database\footnote{\url{http://www.geonames.org/}}.
To identify potential misinformation, we find lists of low-credibility domains in different languages.
For selected countries, we build two networks: retweet and co-sharing, and we apply clustering to find communities.
We then manually label (in two stages) samples of tweets from these communities to identify those in which users are likely to encounter no-vax content.
Finally, we compute several measures to quantify network polarization and information co-sharing, as well as the intensity of cross-national interactions between no-vax communities.
We omit here the simple statistics we ran to compare no-vax communities to others, and examine the users who have been suspended.

\begin{figure}[tbp]
\centering
\includegraphics[width=1\linewidth]{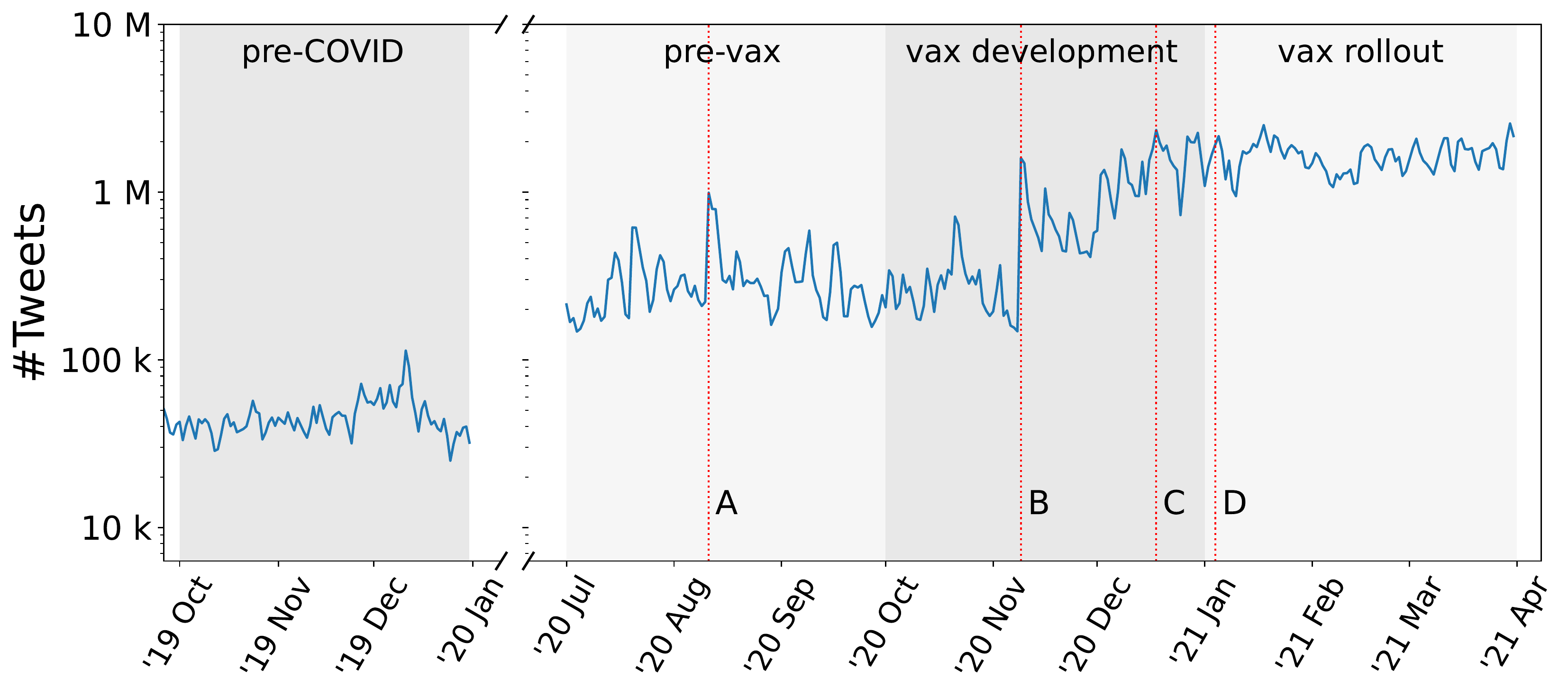}
\caption{\textbf{Volume of the vaccination debate on Twitter.}  Some external events with a significant impact: (A) 2020-8-11: Sputnik V vaccine announced, (B) 2020-11-9: Pfizer-BioNTech vaccine announced, (C) 2020-12-18: Moderna vaccine announced, (D) 2021-1-4: First AstraZeneca vaccine inoculation. 
} 
\label{fig:tweets_volume_tseries}
\end{figure}

\textbf{Dataset.}
We begin by assembling a list of vaccine-related words translated in 18 different languages (\emph{vaccine}, \emph{novax}, \emph{measles}, \emph{MMR}, \emph{vaccinated}, ...), obtaining a set of 459 keywords\footnote{See queries at \url{https://tinyurl.com/344rbjcx}}. 
For each language, we query the Twitter Streaming API\footnote{\url{https://developer.twitter.com/en/docs/tutorials/stream-tweets-in-real-time}} for the tweets containing the keywords in that language (translated by volunteer native speakers) and those in English, applying a language filter.  
For analysis, we choose 4 three-months periods: (1) \emph{pre-COVID} period, from 2019-10-01 to 2019-12-31, (2) \emph{pre-vaccine} period, from 2020-07-01 to 2020-09-30, (3) \emph{vaccine development} period, from 2020-10-01 to 2020-12-31, and (4) \emph{vaccine rollout} period, from 2021-01-01 to 2021-03-31.
See Figure \ref{fig:tweets_volume_tseries} for a summary of daily volume of the dataset.
The volume increased by two orders of magnitude during the pandemic, from 6M tweets in the 3-month pre-COVID period, to 39M in the pre-vaccine period, 91M in the vaccine development period, and 178M in the vaccine rollout period.
To check the completeness of our data, we run an Historical API in pre-COVID period with the same keywords.
Due to the accounts suspension or post removal by the users themselves, a wide fraction of the tweets (72\%) has not been retrieved by the this API, showing that a such a dataset cannot be retrieved by retrospective search.
Moreover, we take advantage of the passage of time in order to re-visit the most notable accounts (present in the networks described below) using the Twitter Get User API call\footnote{\url{https://developer.twitter.com/en/docs/twitter-api/v1/accounts-and-users/follow-search-get-users/api-reference/get-users-show}} to check on their status, noting specifically if the accounts has been suspended by the platform or deleted by the user.

\begin{figure}[!ht] 
\centering
\includegraphics[width=0.95\linewidth]{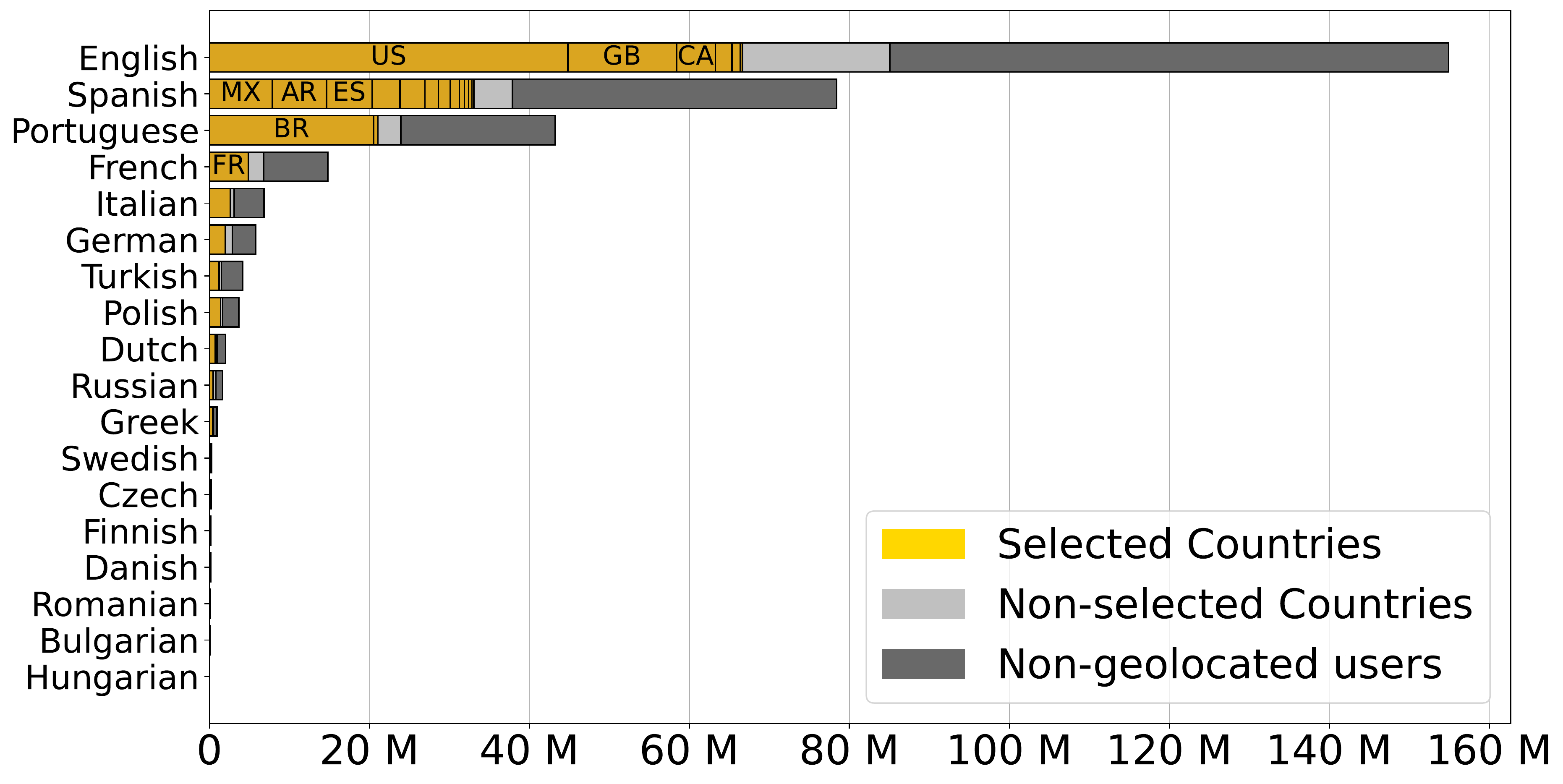}
\caption{\textbf{Tweets volume per language and country.} Number of tweets in the data collection by language and geolocated country (only largest countries labeled).
}
\label{fig:lang_geo_bars}
\end{figure}

\noindent\textbf{Geolocation.}
To capture country-specific dynamics of the social networks, we geolocate the users: we match the location they provide in their description with the geographical database of locations from GeoNames.
Manually verifying matching accuracy, we filter out over 500 words often associated with non-locations in this field. 
To further limit wrong geolocations, (i) we remove the geolocation of users that changed their country locations in the observed period (ii) we manually inspect users responsible of $>$50\% of retweets between two pairs of countries in one period, assuming that a user that is heavily retweeted from another country is more likely to be wrongly geolocated.
Under these conditions, we geolocate 48.7\% of the users.
This then allowed us to select countries for the study (as the focus was on the Western languages, we selected countries from Europe, America and Oceania).
To this end, we filter countries with more than 2000 unique user in each period, obtaining 28 countries\footnote{The complete list (in 2-letter ISO standard): \emph{US, BR, AR, GB, ES, MX, FR, CA, TR, VE, AU, CO, IT, CL, DE, PT, IE, PY, EC, RU, UY, NZ, PL, NL, PE, CU, PA, GR}.}, spanning 11 languages.
See Figure \ref{fig:lang_geo_bars} from Appendix for further details on tweets volume per language.

\noindent\textbf{Low-credibility domains.}
Following previous literature on misinformation tracking~\cite{gallotti2020assessing, pierri2020diffusion}, we collect a list of low-credibility domains.
As sources of low-credible websites we rely on Bufale.net \footnote{\url{https://www.bufale.net/the-black-list-la-lista-nera-del-web/}} (Italian), Wikipedia \footnote{\url{https://en.wikipedia.org/wiki/List_of_fake_news_websites}} (English), Media Bias/Fact-check \footnote{\url{https://mediabiasfactcheck.com/}, conspiracy-pseudoscience and questionable sources} (English), Le Monde.fr \footnote{\url{https://www.lemonde.fr/les-decodeurs/article/2018/10/17/les-fausses-informations-perdent-du-terrain-sur-facebook_5370461_4355770.html}} (French), dwrean.net \footnote{\url{https://www.dwrean.net/2018/12/fake-news-greek-sites.html}} (Greek), obtaining a list of 1732 domains.
The fact that we were unable to find lists for less used languages is an important limitation of this work, which we discuss in the Discussion section.

\noindent\textbf{Network reconstruction.}
For each country, for each period, we build a Retweet (RT) network and a Co-sharing (CO) network. 
In order to limit the number of geolocation mismatches and filter users belonging to debates in other countries, we constrain the tweets considered for each country to the most common language in our data in that country, among the languages spoken in the country (according to Wikipedia). 
The RT network is a directed weighted graph, where each node is a user, and the weight of the directed link $ij$ is the number of times that user $i$ retweeted $j$, 
The CO network is an undirected weighted graph, where each node represents a user, and the weight of the undirected link $ij$ is the number of unique URLs shared by both the users. 
After basic pruning according to \cite{garimella2018quantifying}, an average RT network includes the 65\% of users in the dataset in a country and period, while a CO network the 16\%.
Focusing only on the Giant Connected Component (GCC) of the constructed networks, on average, 92\% of the nodes of the RT, and 76\% of the nodes of the CO networks are in the GCC, while the average Overlap Coefficient (OC), defined as $OC(A, B) = \frac{A \cup B}{min(|A|,|B|)}$, between the sets of users in RT and CO networks increases from 0.72 (pre-COVID period) to 0.86 (vaccine rollout period), as people share more URLs. 

\noindent\textbf{Hierarchical Clustering.}
Next, we apply a community detection algorithm to cluster the users of the RT and CO social networks.
Since the goal is to find a small number of large groups of users, we adopt hierarchical clustering, instead of unsupervised algorithms (e.g. Louvain), that finds the optimal partitioning with a very large number of often small communities.
The clustering process involves the following steps:
\begin{enumerate}
    \item Build the dendrogram of the hierarchical clustering using Paris algorithm \cite{bonald2018hierarchical};
    \item Compare the partitions obtained with cutoffs at heights 2, 3, 4 and 5 (that is, having this number of communities);
    \item Pick the partition with the highest modularity;
    \item If more than 90\% of nodes are in the same community, compare the partitions with cutoffs at the following 5 heights and repeat from step 3.
\end{enumerate}

With this procedure, we ensure that 90\% of the users are partitioned into at least two communities, but no more than five (while it is possible to have many small communities, that comprise less than 10\% of nodes).
Using this method, we find on average 6.7 communities in RT networks, and 5.0 communities in CO networks, with a maximum of 20.

\noindent\textbf{Labelling.} 
To identify communities where users are exposed to no-vax content, we label a sample of tweets shared in each community. 
First, we filter out small communities, by considering only those with more than 1\% of the users of the network, resulting in 400 communities.
Next, we randomly sample 20 tweets from each community, making a total of 8000 tweets. 
Twelve people were involved in the labelling, all of them with a background in vaccine debate and knowledgeable of the language used in the tweet to label.
Furthermore, we translate all tweets into English using Google Translate to allow for cross-check.
Each person labelled between 600 and 1000 tweets, with an overlap of 20 tweets with other annotators.
The tweets are labelled as ``pro-vax", ``no-vax" or ``other".
We labelled as pro- or no-vax only the tweets that were clearly supporting or discrediting vaccines, respectively. 
For this reason, more than half of the labels are ``other'', comprising non-relevant posts, unclear positions, discussion on other policies, and all generic pieces of news that do not express a stance. The task of distinguishing between pro- and anti-vaccination stances proved to be fairly easy, with a Cohen's kappa computed on a overlapping set at $k$ = 0.84 (only 3\% received different labels).
Instead, the task involving the ``other'' label proved to be more difficult, with $k$ = 0.51 for the 3-class setting (disagreement of 26\%), much of it due to the confusion between ``other'' and ``pro-vax'' (disagreement of 20\%). 
However, note that we are only interested in distinguishing between anti-vaccination stances and the rest.

\begin{figure}[!ht]
\centering
\includegraphics[width=0.65\linewidth]{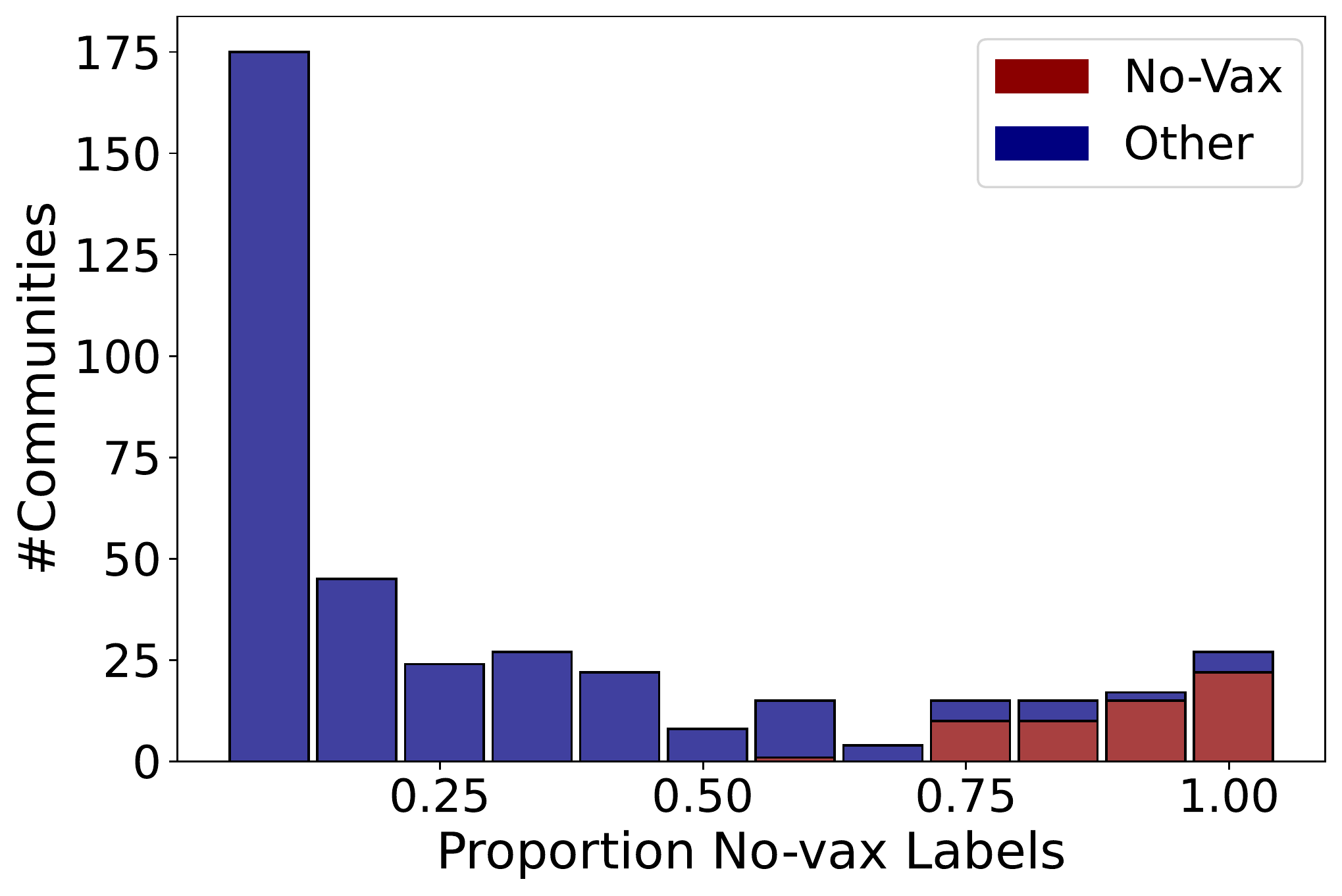}
\caption{\textbf{Number of communities with a certain proportion of tweets labelled as no-vax in the first labelling.} 
In red communities with more than 10 anti-vax labels in the two rounds (no-vax communities).
}
\label{fig:hist_prop_novax}
\end{figure}

To improve the quality of the labels, we then proceeded to a second round of annotation, focusing on the communities which have a majority of content with a no-vax stance.
Specifically, we choose the communities with a majority of the no-vax tweets, and annotate the 10 most popular tweets in each (excluding the 50 most popular in the whole network). 
The second labeling stage encompassed 82 communities, totalling in 820 tweets.
At this stage, the Cohen's kappa considering the 3 classes is $k$ = 0.64. 
Finally, we define a community as no-vax if the total number of `no-vax' labels in the two rounds is greater than 10, resulting in 58 communities. 
Since some networks have more than one anti-vax community, we have 52 networks with a no-vax community i.e. a community where users are substantially exposed to no-vax content (see Figure \ref{fig:hist_prop_novax}, Appendix). 

\noindent\textbf{Random Walk Controversy.}
Following previous literature \cite{garimella2016quantifying, garimella2018political, cossard2020falling}, we use the Random Walk Controversy (RWC) score to quantify the polarization between the communities labeled as no-vax and the rest of the network. 
Given a RT network, partitioned in two clusters $X$ and $Y$, RWC is defined as $RWC = P_{XX} P_{YY} - P_{XY} P_{YX}$, where $P_{AB}=P(\text{a random walk ended in B started in A})$.
Intuitively, it represents the difference in probability for an average user in the network to be exposed to information from their own side versus from the opposing side. 
Spanning $[0,1]$, an RWC close to 1 represents a polarized social network with two distinct groups that do not endorse each other’s opinion, while an RWC close to 0 represents a non-controversial topic where both opinions are equally likely to be received.

\noindent\textbf{Normalized Mutual Information.}
We quantify the echo-chamber effect by measuring the extent to which users from different retweet communities share the same sources of information, as a proxy for the information siloing in an echo-chamber.
To do this, we employ Normalized Mutual Information (NMI) to gauge the similarity between the RT and CO communities \cite{danon2005comparing}, by using the \texttt{normalized$\_$mutual$\_$info$\_$score} module in the Python package \texttt{scikit-learn}.
Spanning $[0,1]$, a NMI of 1 means that the community structure is the same between the two networks, while a low NMI means different community structures. 
Note that this metric does not use the opinion leaning determined in the labeling step, and thus can be computed for any country/period network, whether or not it has a no-vax community.

\noindent\textbf{Normalized Retweets Volume.}
To assess the extent to which one country retweets another, we compute a normalized retweeting volume for each pair of countries.
To this aim, Figure \ref{fig:heatmap_cross_rt} (a) shows $n_{ij}$, the number of retweets from one country $i$ to another country $j$ ($a_{ij}$), divided by the total number of RTs by users in country $i$ ($s^{out}_i$) and the total number of RTs to users in country $j$ ($s^{in}_j$), multiplied by the total number of RTs by all countries, 

\begin{equation*}
n_{ij} = 
\begin{cases}
\frac{a_{ij}}{s^{out}_i s^{in}_j} E & \text{if } i \neq j \\
0 & \text{if } i = j
\end{cases}
\end{equation*}

where $E=\sum_i s^{out}_i =\sum_i s^{in}_i$. Note that $s^{out}_i s^{in}_j / E$ is the expected number of RTs from country $i$ to country $j$ in the random graph with same nodes strengths.
Hence, we have $n_{ij}$ $>$ 1 if $i$ retweets $j$ more than in a random baseline context.
Since the vast majority of retweets are within the same countries, $a_{ij} \gg s^{out}_i s^{in}_j / E$ if $i \neq j$, otherwise $a_{ij} \ll s^{out}_i s^{in}_j / E$.
In order to highlight cross-border interactions, in the plots we show only off-diagonal elements.

\noindent\textbf{Cross-border Interactions between No-vax Communities.}
We measure the strength of ties between users in no-vax communities in different countries by comparing the number of retweets between these users with the retweets between the rest of users in the same countries. 
In particular, we define $V^K_i$ as the set of users in communities with stance $K$ in a country $i$, where $K$ can be A (anti-vax) or O (others).
Let us define $E^K_{ij}$ as the retweets from users in $V^K_i$ to users in another country $j$ with the same stance $K$, $V^K_j$.
One can thus measure the density $\delta_{ij}^{K} = E^K_{ij} / {V^K_i \,V^K_j}$, the ratio of observed retweets and the total possible pairs between sets $V^K_i$ and $V^K_j$, that is the probability that two random users respectively in $V^K_i$ and $V^K_j$ are connected.
Figure \ref{fig:heatmap_cross_rt} (c) show the ratios $\theta_{ij} = {\delta^A_{ij}}/{\delta^O_{ij}}$ of the densities between users belonging to no-vax communities $\delta^A_{ij}$ in countries $i$ and $j$ and the rest of users in the same countries $\delta^O_{ij}$.
If $\theta_{ij} = {\delta^A_{ij}}/{\delta^O_{ij}} > 1$, it means that the probability that two random users in no-vax communities in countries $i$ and $j$ are connected is higher with respect to two random users in rest of network of the same countries.

\begin{figure*}[tb] 
\centering
\includegraphics[width=0.95\linewidth]{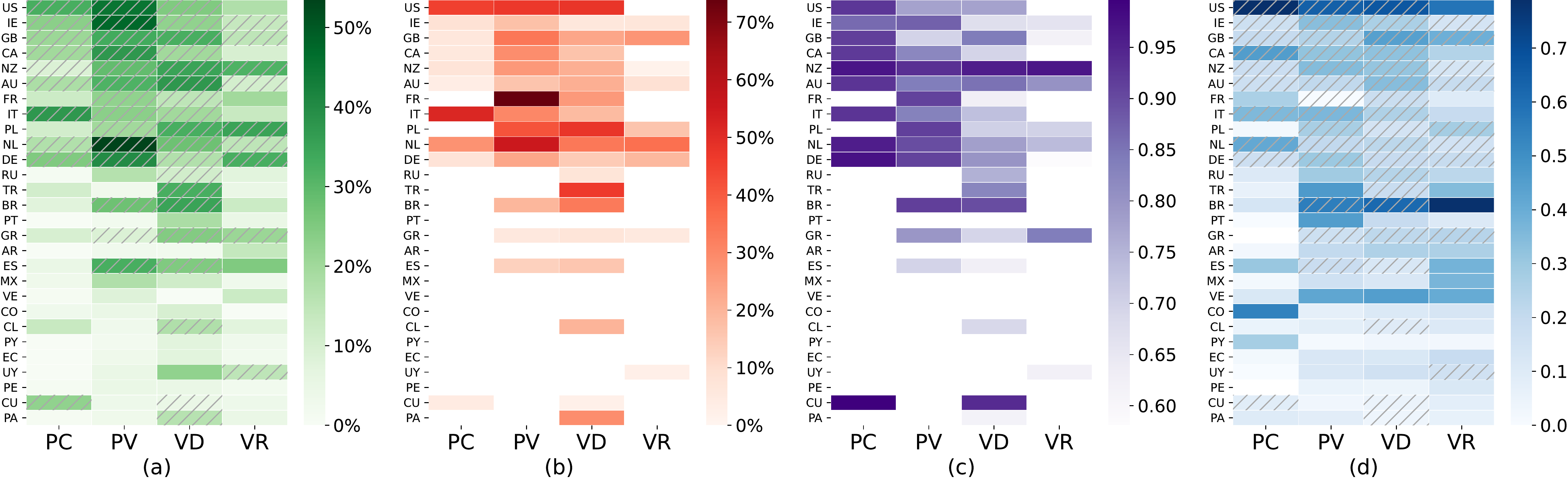}
\caption{\textbf{Characterization of no-vax communities for each country and period considered.}
(a) Proportion of tweets labelled as no-vax, (b) proportion of users in no-vax communities with respect to the size of the RT networks (average in the 4 periods: 16.9$\%$, 30.9$\%$, 23.1$\%$, and 13.7$\%$), (c) RWC between no-vax communities and the rest of the networks (average in the 4 periods: 0.94, 0.84, 0.76, 0.73), (d) NMI between retweets and co-sharing communities.
Countries with no-vax communities are marked with dashed lines.
Periods acronyms refer to pre-COVID (PC), pre-vaccines (PV), vaccine development (VD), vaccine rollout (VR).}
\label{fig:heatmaps_countries_periods}
\end{figure*}

\section{Results}

\noindent \textbf{Polarization of the vaccination debate.}
We begin by examining different measures of polarization and no-vax activity in different countries in the four time periods.
Figure \ref{fig:heatmaps_countries_periods} (a) shows that a high presence of no-vax tweets in a certain country and period is often associated with the presence of a community labelled as no-vax (dashed lines). 
This implies that no-vax content is generally clustered and not homogeneously distributed in the RT network, suggesting that the debate is polarized, as we will show below.
Further, we find that no-vax communities are generally present in English-speaking countries (e.g. with respect to Spanish speaking ones).
However, some of the relatively largest country-specific no-vax communities appear in France, Italy, Netherlands, Poland, and the United States (Figure \ref{fig:heatmaps_countries_periods} (b)).
No-vax communities are particularly present in pre-vaccine and vaccine development periods, where they also span a larger fraction of users with respect to other periods.

Turning to potential echo-chambers in these networks, 
we find that the Random Walk Controversy (RWC) score is overall very high (Figure \ref{fig:heatmaps_countries_periods} (c)), indicating that the vaccination debate is generally highly polarized.
However, it decreases substantially over time, suggesting that users in no-vax communities became less isolated in the vaccination discourse during the COVID pandemic.
Secondly, we investigate whether the users in the no-vax communities are exposed to information sources that are different from the rest of users by considering Normalized Mutual Information (NMI).
Despite the NMI being independent of the labelling of the communities, 
Figure \ref{fig:heatmaps_countries_periods} (d) shows that, on average, the NMI of the networks with a no-vax community is higher than the others (0.27 vs 0.22,  $p$ $<$ 0.05), indicating that users in no-vax communities tend to have common information sources.
Some countries, such as the U.S. and Brazil, show a especially high NMI, indicating that the polarization in the retweet network is reflected in the different content shared.
Spanish speaking countries, conversely, are less polarized than English speaking countries (average 0.15 vs. 0.33, $p$ $<$ 0.001).


\begin{figure}[tbp]
    \centering
    \includegraphics[width = \linewidth]{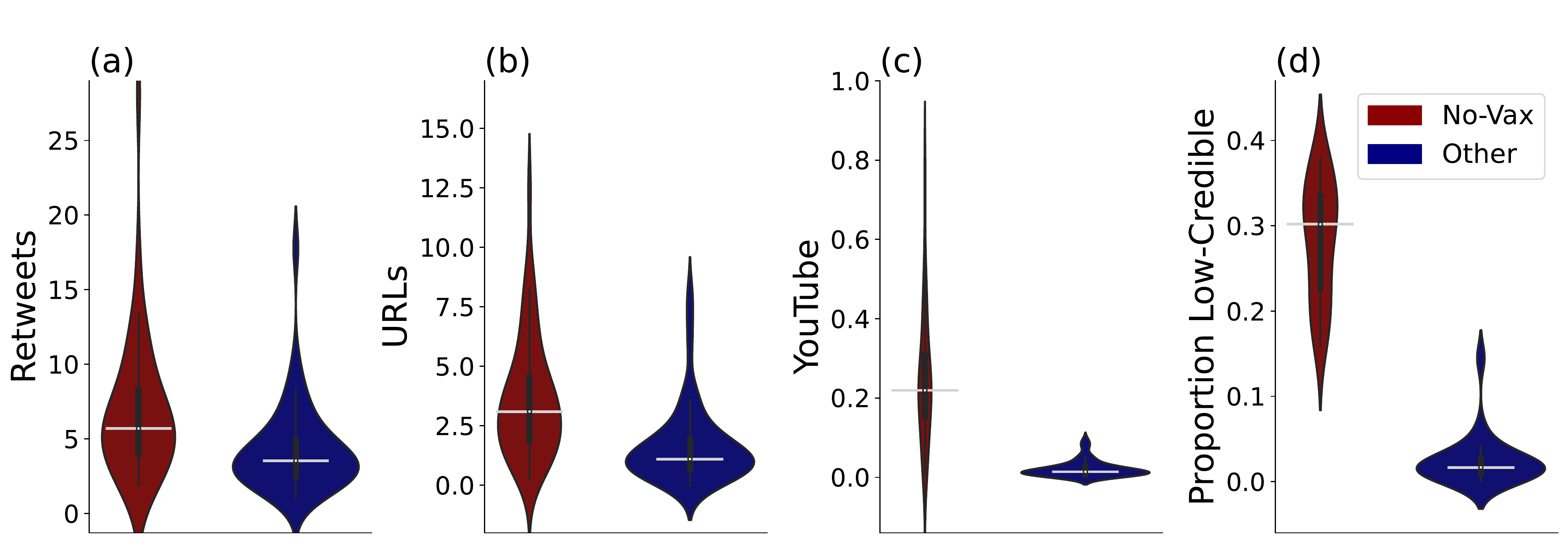}
    \caption{\textbf{Behavior of users in no-vax communities vs other users.} (a) Average retweets, (b) average URLs, (c) average YouTube URLs, and (d) proportion of low-credible domains, shared by users. Note that low-credible domains are collected only in Italian, French, English, Greek, hence the plots refer to countries speaking these languages.}
    \label{fig:violins}
\end{figure}
\noindent\textbf{Characterizing users in no-vax communities.}
Considering the behavior of users in no-vax communities, we find that they are more likely to retweet (Figure \ref{fig:violins}(a)), share URLs (Figure \ref{fig:violins}(b)), and especially URLs to YouTube (Figure \ref{fig:violins}(c)) than other users.
Furthermore, the URLs they post are much more likely to be from low-credible domains (Figure \ref{fig:violins}(d)), compared to those posted in the rest of the networks. 
The difference is remarkable: 26.0\% of domains shared in no-vax communities come from lists of known low-credible domains, versus only 2.4\% of those cited by other users ($p$ $<$ 0.001).
The most common low-credible websites among the no-vax communities are \emph{zerohedge.com}, \emph{lifesitenews.com}, \emph{dailymail.co.uk} (considered right-biased and questionably sourced) and \emph{childrenshealthdefense.com} (conspiracy/pseudoscience).
These findings extend to the existing literature on English-language vaccination rhetoric to multi-lingual, international scope by confirming the elevated social engagement in anti-vaccination communities \cite{germani2021anti}, and provide additional evidence of the misleading nature of the popular COVID-related YouTube videos \cite{dutta2020youtube}.

\begin{figure*}[tbp]
    \centering
    \includegraphics[width = 0.95\linewidth]{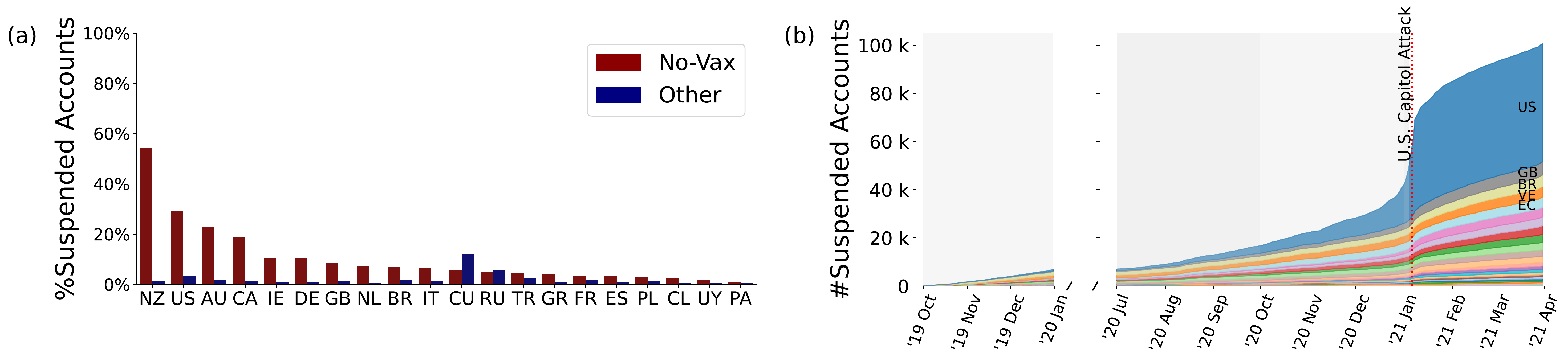}
    \caption{\textbf{Suspended users per country in no-vax communities.} (a) Average proportion of suspended accounts per country, in the periods in which no-vax community has been detected, computed separately for no-vax side and rest of users. 
    (b) Number of suspended accounts as a function of the date they posted their last tweet, coloured by country. }
    \label{fig:hist_susp}
\end{figure*}
Next, we investigate the effects of content moderation by Twitter on the vaccination debate.
We find that the average proportion of suspended accounts in no-vax communities is much larger than the rest of users, for each country and period considered (average 13.3\% vs 1.8\%, $p$ $<$ 0.001) (Figure \ref{fig:hist_susp} (e)).
The highest proportions of suspended accounts are found in English speaking countries, Germany and the Netherlands---those which also show a larger presence of no-vax content, compared to other countries.
Further, a large portion of suspensions come after the January 2021 U.S. Capitol attack in Washington, D.C.\footnote{The suspensions were announced by Twitter \url{https://blog.twitter.com/en_us/topics/company/2021/protecting--the-conversation-following-the-riots-in-washington--}} (Figure \ref{fig:hist_susp} (f)).
The proportion of suspended accounts that come from the U.S. jumped from 38\% before January 1st, to 77\% during the days around the Washington riots (January 1--12).
Note that (i) $89\%$ of U.S. users suspended belong to the no-vax community in the vaccine development period,
(ii) the user \texttt{realDonaldTrump} (suspended on January 8th) is one of the most popular accounts in the no-vax communities of the first three periods, and 
(iii) in the last period, a no-vax community is not present in the U.S. RT network, indicating that the suspension of U.S. accounts following the Washington riots heavily impacted on the vaccination debate.
These findings suggest that political leaning is often associated with strong stances taken in the vaccination debate (in line with previous literature \cite{cossard2020falling, motta2020right}) and that actions taken in the political domain may greatly impact the quality of the public health discourse.

\begin{figure*}[tbp]
\centering
\includegraphics[width=1\linewidth]{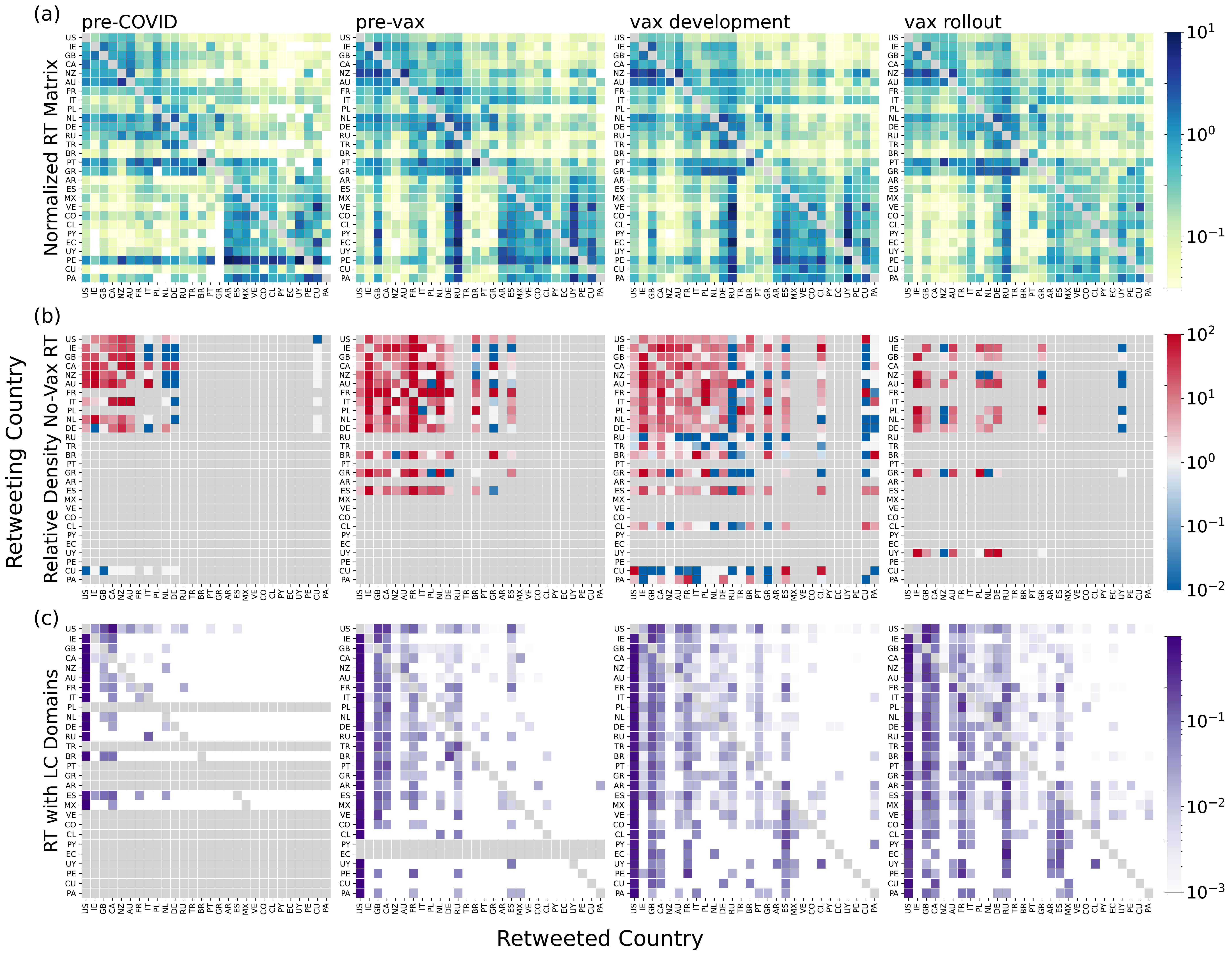}
\caption{
\textbf{Cross-border information flows in the global vaccination debate.}
(a) Normalized number of retweets (excluding diagonal elements from the plot, colored in grey), (b) Probability of interaction between users in no-vax communities from one country to another, with respect to the interactions between other users from the same pair of countries (see Methods). Darker red (blue) elements of the matrices represent higher (lower) tendency of cross-border interactions between users in no-vax communities with respect to other users (countries without no-vax communities colored in grey), (c) Proportion of URLs that come from Retweeted Country among the low-credible domains imported by Retweeting Country (countries importing less than 10 low-credible URLs are coloured in grey).
Element $a_{ij}$ of each matrix represents information flow from country $j$ to country $i$.
}
\label{fig:heatmap_cross_rt}
\end{figure*}

\noindent\textbf{Cross-border information spillover in the global vaccination debate.}
Next, we quantify the information spillover across countries by considering the number of retweets from one country to another, normalized by the total number of retweets produced and received in the two countries (Figure \ref{fig:heatmap_cross_rt} (a)).
First, one can observe language homophily, indicated by darker regions in the top left (English) and bottom right (Spanish) of the panels, as well as the pair Portugal-Brazil, in all periods.
The darker patches corresponding to interactions between Germany and the Netherlands, and Germany and Turkey, also reflect possible cultural or expat relationships.
Second, the cross-border interaction matrices are not symmetric: information generally flows with a preferred direction. 
For instance, Spanish-speaking countries retweet English-speaking ones much more than the opposite. 
Note that the United States is central in the global information flow (despite flows being normalized), being a net exporter of information to the rest of the world comparing in-flows vs out-flows of information for each country). 
Interestingly, from pre-vax period, Russia is also a net exporter, especially to South American countries: some of the most used hashtags in pre-vaccine and vax development periods are \#sputnikesesperanza and \#sputnikparaelpueblo.

In Figure \ref{fig:heatmap_cross_rt} (b), we quantify the strength of cross-border interactions between users in no-vax communities with respect to the rest of users.
We find that cross-border interactions between users in no-vax communities are generally much stronger, sometimes by orders of magnitude, than interaction from the rest of users, creating a tightly-knit global no-vax network. 
In particular, users in no-vax communities of English-speaking countries, Germany, and the Netherlands are tightly connected in all periods. 
Conversely, users in no-vax communities from Cuba and Russia are quite isolated (adding to their unusual user suspension statistics).
Again, cross-border interactions can be asymmetric: for instance, in pre-COVID period, users in no-vax communities in Germany and the Netherlands retweet users from other countries but not vice versa.

Finally, we focus on misinformation flows across countries by considering the fractions of low-credible domains imported per country (Figure \ref{fig:heatmap_cross_rt} (c)), that is, the fraction of tweets pointing to low-credible URLs, over the total number of retweets, from one country to another.
We stress that we consider flows of low-credible information across borders spread by both humans or bots, without engaging in the difficult task of distinguishing them, as we are interested in quantifying how exposed is a certain country A to misinformation coming from country B.   
As in the previous case, the matrices show a clear asymmetry.
U.S. users are responsible for exporting a large fraction of misinformation 
to the rest of the world: 
68\% of all low-credible URLs retweeted worldwide come from U.S. (average over the four periods), a proportion much higher than the total volume (42\%) retweeted from U.S..

\begin{figure}[!ht]
    \centering
    \includegraphics[width = 0.9\linewidth]{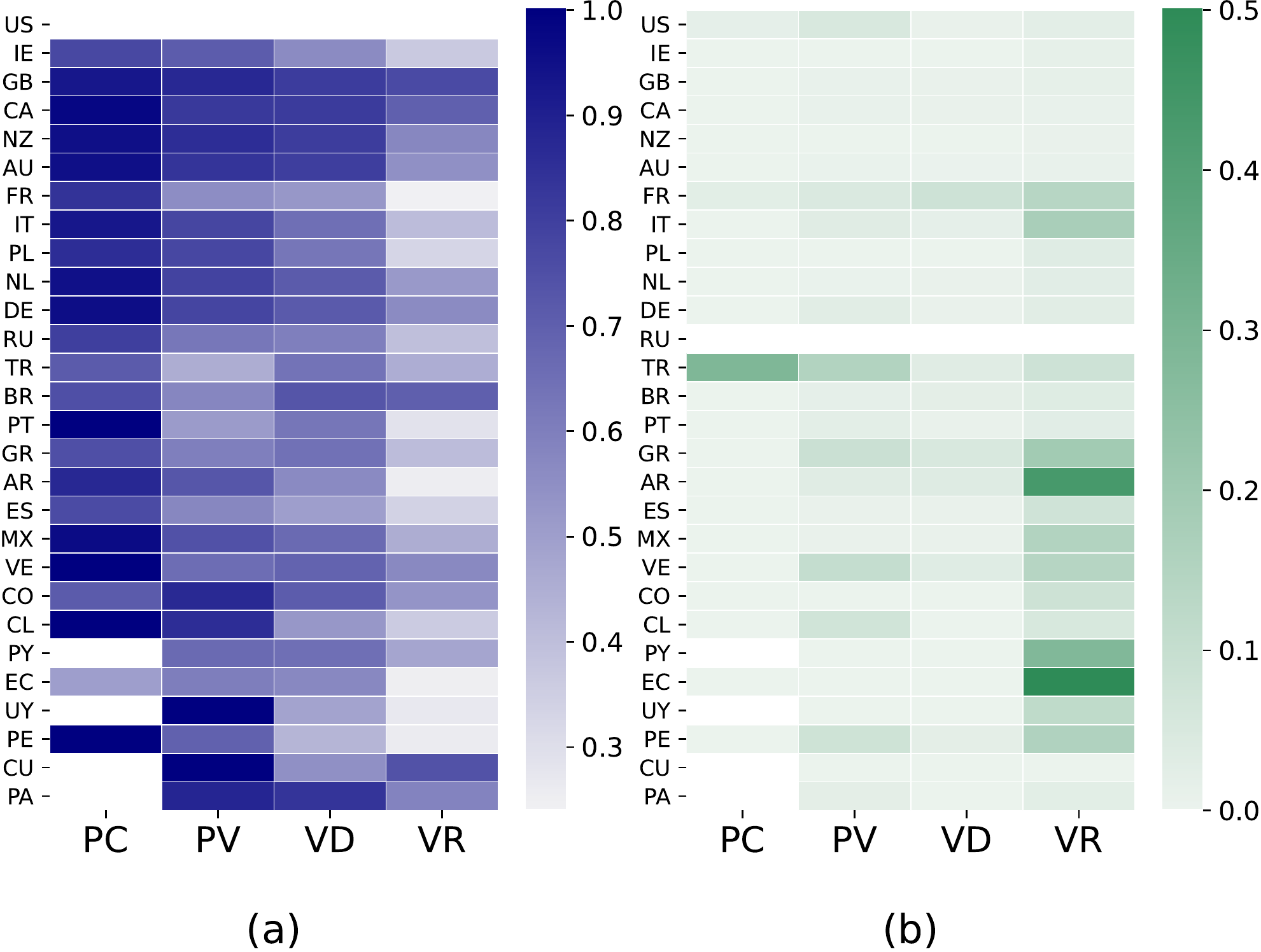}
    \caption{{\textbf{URLs with low-credible domains exported by U.S. and Russia.} 
    (a) Proportion of low-credible URLs imported from U.S., for each country and period considered. 
    Overall, the proportion of URLs that comes from U.S. is 93\%, 79\%, 74\%, 56\%. 
    (b) Proportion of URLs imported from Russia, for each country and period considered. 
    In general, countries import 1\%, 2\%, 1\%, 10\% of the low-credible domains from Russia.
    }
    }
    \label{fig:LC_from_USRU}
\end{figure}

Interestingly, the fraction of low-credible URLs coming from U.S. dropped from 74\% in the vax development period to 55\% in the vax rollout. 
This large decrease can be directly ascribed to Twitter's moderation policy: 46\% of cross-border retweets of U.S. users linking to low-credible websites in the vax development period came from accounts that have been suspended following the U.S. Capitol attack (see Figure \ref{fig:LC_from_USRU} (a), Appendix).
Notice that Twitter's account purge significantly impacted misinformation spread worldwide: the proportion of low-credible domains in URLs retweeted from U.S. dropped from 14\% to 7\%.
Finally, despite not having a list of low-credible domains in Russian, Russia is central in exporting potential misinformation in the vax rollout period, especially to Latin American countries.
In these countries, the proportion of low-credible URLs coming from Russia increased from 1\% in vax development to 18\% in vax rollout periods (see Figure \ref{fig:LC_from_USRU} (b), Appendix).

\section{Discussion} 

The international, multi-lingual nature of the data we present here supports the ongoing efforts in monitoring non-English debate around the topic of vaccination \cite{di2022vaccineu, pierri2021vaccinitaly, haouari2021arcov}.
Using it, we reveal the increasingly globalized nature of the vaccination debate, as the COVID-19 vaccines were proposed, developed, and deployed. 
This increased globalization had a marked impact on the vaccine hesitant discourse: not only did the prominence of the no-vax communities increase within their countries, but their cross-border connections strengthened around the world.
We show that users in these communities are much more prone to sharing potential misinformation than other users, also across national borders.

Further, the real-time nature of the data collection allowed us to capture Twitter's content moderation efforts, which proved to be uneven both across countries and time.
The users blocked immediately following the January 6th Washington riots were responsible for a substantial amount of misinformation spread -- both within the United States, and, crucially, internationally.
Thus, we paint a picture of a ``global no-vax Twitter network'' that calls for international collaboration of both public health and technology experts.

There are several important limitations in our study.
First and foremost, it is well known that Twitter users are not representative sample of the real population, but biased toward more educated, urban, younger and male individuals \cite{pew2021social}. 
Furthermore, Twitter usage wildly differs among the countries under consideration, so that cross-country comparison should be taken with caution. 
However, to the best of our knowledge there are no means to collect real-time, representative data at this spatial and temporal scale.
Note that we do not engage in bot detection, as this task is notoriously difficult \cite{martini2021bot}, and most importantly, misinformation can be spread by complex interactions by bots and humans \cite{gallotti2020assessing}.
Moreover, our study is limited to 11 chosen languages and to the 4 languages for which low-credible domains were collected, a limitation necessary in order to control the cultural heterogeneity of the data analyzed.
Still, the fact that low-credible domains in other languages were not found (for instance, it was challenging to find a reputable list of low or high-credible domains in Russian) means that the potential misinformation flows presented here is a lower bound -- one which should be expanded using additional resources.

Future work should also be devoted to include countries from Africa and Asia, as well as to update and extend to other languages the list of low-credible information sources.
For this latter task, one could leverage the identification of no-vax communities -- more susceptible to share low-credible information -- proposed in the current study. 
Other possible limitations of this work include the method used to identify no-vax communities, hierarchical clustering of the RT network and labelling of popular tweets in the resulting communities, which may be sensitive to the thresholds adopted.
Note, however, that these methods are not aimed at detecting the stance of single users about vaccination, but to identify large clusters of users exposed to a certain kind of anti-vaccine narrative.

\subsection{Broader Impact}

Despite the platform's tweet flagging and removal policies around COVID-19 \footnote{\url{https://help.twitter.com/en/rules-and-policies/medical-misinformation-policy}}, it is the bout of account suspensions around the Washington riots that made the most impact on the national and international spread of vaccine-related misinformation, suggesting that the political concerns elicit much stronger curbing of the freedom of speech than the health one.
More documentation of the causal link between online misinformation and adverse health outcomes may provide a more solid ground for making censorship decisions for both the platforms and the politicians governing them.
For instance, a randomized controlled trial in the UK and the USA showed that ``relative to factual information, recent misinformation induced a decline in intent of 6.2 percentage points''~\cite{loomba2021measuring}.
Further, CDC and the Kaiser Family Foundation estimate that the lack of action early in the pandemic may have contributed to deaths of hundreds of thousands by June 2021~\cite{amin2022mortality}. 

Further, this study illustrates the impact of one social media platform's editorial policies on the international public health discourse, especially when the country involved is as culturally influential as the United States. 
Without examining in detail the content shared by the suspended accounts, we cannot be certain that the accounts indeed were sharing harmful content.
Monitoring censorship activities of major platforms (either triggered by internal policies or governments' requests) is important for assessing the users' freedom of speech. 
For instance, Electronic Frontier Foundation has recently criticized social media platforms for blocking political dissidents who a decade ago used the same platforms to ``push for political change and social justice''\footnote{\url{https://www.eff.org/deeplinks/2020/12/decade-after-arab-spring-platforms-have-turned-their-backs-critical-voices-middle}}. 
Fortunately, ``de-platforming as censorship'' is a topic of ongoing deliberation at the EU's Internet Governance Forum, involving civil society and governmental representatives\footnote{\url{https://www.intgovforum.org/en/content/igf-2022-ws-52-de-platforming-as-censorship-means-in-the-digital-era}}. 

International perspective may also benefit the tracking of malicious actors, such as semi- or fully-automated accounts, networks of colluding agents, and sources of poor quality content. 
Before the pandemic, it has been shown that accounts identified as Russian trolls were more likely to tweet about vaccination~\cite{broniatowski2018weaponized}.
During the pandemic, Russian trolls often posted misinformation concerning personal dangers of vaccines, purported civil liberty violations, and vaccine conspiracies~\cite{warner2022vaccine}.
Since the beginning of the Ukraine war, it has been noted that anti-vaccine content has diminished dramatically, potentially due to the additional blocking of Twitter in Russia, and re-focusing of the conspiratorial attention on Ukraine\footnote{\url{https://www.theguardian.com/media/2022/mar/04/bot-holiday-covid-misinformation-ukraine-social-media}}.
Our findings suggest that changes in governance and censorship may encourage or discourage the flow of potential misinformation from the states with known affinities.


\subsection{Ethical Considerations}

Although the data was collected using the platform's own API, resulting only in posts which were posted publicly, it is possible that some users were not aware of the scope of their potential audience. 
Thus, we follow the platform's Terms of Service, and share only the IDs of the tweets, so that when the data is re-collected, deleted content will not be available (notably limiting the reproducibility of any Twitter-based study).
Further, the multinational nature of the data captures wildly varying biases in the way people around the world are able to access internet, or Twitter specifically. 
Local barriers to access to the internet and local blocks of the platform itself shape the communities captured in the present study. 
For instance, dissidents or those wishing to remain anonymous will likely not share their location information on their profile, and would not be captured as being a part of a country's discussion.
Moreover, the data may capture vulnerable groups, including those experiencing or who are at risk of specific health conditions, those having financial barriers to healthcare, and even those more susceptible to misinformation, among others. 
Despite the negative connotation around ``no-vax'' communities, users found to propagate harmful information may first and foremost endanger themselves by following faulty advice.
Thus, we would discourage the future researchers from publishing verbatim tweet text in order to preserve user privacy.

Finally, in this work we present tools that may be used to track and profile groups of Twitter users around a topic, which may be then used by both the platform and the government.
However, such tools may also be used to target communities for harassment, doxing (providing private user information in aims to harm or intimidate the person), and other abuses. 
On one hand, it is the responsibility of the platforms and their communities to uphold civil code of conduct and block the abusers. On the other, we call for the research community, as well as corporate and governmental actors, to use these tools ethically, with minimal harm to the subjects.


\subsection{Acknowledgements}

The authors acknowledge support from the Lagrange Project of the Institute for Scientific Interchange Foundation (ISI Foundation) funded by Fondazione Cassa di Risparmio di Torino (Fondazione CRT). 
This work has also received funding from the European Union’s Horizon 2020 research and innovation programme - project EpiPose (Grant agreement number 101003688) and project PANDEM-2 (Grant agreement number 883285). This work reflects only the authors’ view. The European Commission is not responsible for any use that may be made of the information it contains.
Most of this work was performed while M.S. and A.P. were at ISI Foundation.

\subsection{Data \& Code availability}
All codes generated in the project have been deposited to GitHub (\url{https://github.com/jaclenti/global-vax-misinfo}).

All data collected in the project have been deposited to Zenodo (\url{https://zenodo.org/deposit/7113013}).

\subsection{Author contributions}
All authors contributed to developing the main ideas of this study.
J.L. analysed the data and prepared the figures. 
M.T. contributed in the data collection.
Y.M. and M.S. led the interpretation of the results and the preparation of the manuscript, with contributions from all authors. 
All authors critically reviewed and approved the final version of the manuscript.

\subsection{Competing interests}
The authors declare no competing interests.

\section{Appendix - Additional Figures}

\FloatBarrier

Below, additional figures elaborate on the data selection, annotation, and analysis.


\FloatBarrier


\bibliographystyle{abbrv}
\bibliography{bibliography}

\end{document}